# Image Inpainting Based on a Novel Criminisi Algorithm

SONG Yuheng, YAN Hao*

(College of Software Engineering, Sichuan University, Chengdu Sichuan, China)

**Abstract:** In view of the problem of image inpainting error continuation and the deviation of finding best match block, an improved Criminisi algorithm is proposed. The improvement was mainly embodied in two aspects. In the repairing order aspect, we redefine the calculation formula of the priority. In order to solve the problem of error continuation caused by local confidence item updating, the mean value of Manhattan distance is used for replace the confidence item. In the matching strategy aspect, finding the best match block not only depend on the difference of the two pixels, but also consider the matching region. Therefore, Euclidean distance is introduced. Experiments confirm that the improved algorithm can overcome the insufficiencies of the original algorithm. The repairing effect has been improved, and the results have a better visual appearance.

## 0 Introduction

Image inpainting is an important part of image restoration research. Its purpose is to use the neighborhood information of the damaged area of the image to fill and repair the damaged area according to a certain order and rules, so that the restored image and the original image can be consistent with good visual effects. Image restoration technology has been widely used in image scratch reparation, IBR (image-based rendering), cultural relic protection, digital special effects and etc. In recent years, it has become one of the hotspots in the field of image processing. Image restoration is a morbid problem, and currently there is no good and clear solution. [1]

Image restoration technology is mainly divided into two categories. One is the structure-based repair technology. This method uses the edge information of the area to be repaired, and uses a coarse-to-fine method to calculate the direction of the iso-illuminance line. The mechanism spreads the information to the damaged area for better image restoration. The BSCB model, CDD model, TV model, and Euler's elastic model all belong to this method. This type of repair technology has good adaptability for small-size defect repair, but it has a poor effect on large-size defect repair. Another method is based on texture repair technology. The main idea of this method is to select a pixel of the boundary of the damaged area, center the point, and select the texture block of the appropriate size according to the texture features of the image. Find the best matching block, and replace the texture block with this block to complete the image restoration. The Criminisi algorithm proposed in [2] is a typical algorithm in this kind of repair technology. This algorithm can be applied to large-area damaged image restoration and can achieve better visual effects. However, the algorithm still has many shortcomings. Due to the priority calculation formula and the matching principle, it is easy to cause image repair errors and repair discontinuous situation. Many improved algorithms have also been derived from this. In [3], for the problem of failure of geometric structure information in priority calculation, the step evaluation formula separates the confidence item from the structural data item. As a result, the texture information and structural information recovery of the image are improved. In [4] and [5], the degree of influence of the confidence items and data items on the image in the priority calculation is determined. By adding the proportion to determine the priority, the repair order of the image is improved to obtain a suitable repair image. In [6] and [7], the problem that the structural information item is zero causes the priority calculation formula to be invalid. After the threshold separation method is adopted, the priority can still play a role in the case where the structural information item is zero. In [8] and [9], adding regularization factor smoothing priority to confidence items and data items, and change the continuation of repair order to improve the quality of image restoration. In [10], the sample block size is adaptively adjusted by the amplitude variation of the domain gradient vector, which improves the image restoration effect. The [11] defines the pattern calculation formula so that the texture boundary transition tends to be smooth in the case of ensuring the consistency of the image structure. In [12], it introduces the concept of edge enhancement, which improves the phenomenon of continuous optimization of strong edge information and reduces the continuation of matching errors.

In this paper, the Criminisi algorithm is improved by introducing the Manhattan distance to redefine priority calculation formula, and the occurrence of matching error continuous phenomenon is reduced. At the same time, the Euclidean distance is introduced in the matching strategy, and the local information of the image is fully considered to reduce the match error caused by global search.

## 1 Criminisi Algorithm

### 1.1 Criminisi algorithm working principle

The Criminisi algorithm is actually an image restoration algorithm based on structure and texture and it's working principle is shown in Figure 1. $\Omega$ represents the target area to be repaired. $\Phi$ represents a known sample area. $\partial\Omega$ represents the boundary between the known sample area and the target area to be repaired. P is the target pixel to be repaired on the boundary. $\nabla I_P$ is the tangential direction of the iso-illuminance line of the pixel point P. $n_p$ is the normal vector of the tangent of the P point on the boundary between the sample area and the target area to be repaired. $\varphi_p$ is the target block to be repaired centered on point P.

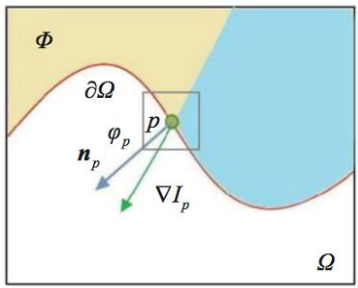

Fig. 1  The principle of the Criminisi algorithm

The priority calculation formula of the Criminisi algorithm is as follows:

$$P(p) = C(p) * D(p) \quad (1)$$

In the above formula, C(p) is a confidence term and D(p) is a data item. They are defined by the following formulas:

$$C(p) = \frac{\sum_{q \in \varphi_p \cap \phi} C(q)}{|\varphi_p|} \quad (2)$$

$$D(p) = \frac{|\nabla I_P^\perp \cdot n_p|}{\alpha} \quad (3)$$

In the above formula, $|\varphi_p|$ is the area of $\varphi_p$. $\alpha$ is the normalization factor, and $n_p$ is the normal vector of point P. When the image is initialized, C(p) is defined as:

$$C(p) = \begin{cases} 0 & \forall_P \in \Omega \\ 1 & \forall_P \in \Phi \end{cases} \quad (4)$$

The above formula indicates that if the point P is within the target area $\Omega$ to be repaired, the confidence C(p) is zero. If the point P is in the known sample area $\Phi$, the confidence C(p) is 1.

### 1.2 Criminisi algorithm step

#### 1.2.1 Priority calculation

First, the image matching block size is determined, and then the boundary of the repair target area is positioned, that is, the $\partial\Omega$ area indicated in the above figure is found and the confidence calculation of all points on the boundary is performed as shown in Equation 2 and 3. The calculation of the gradient data item is performed to obtain the priority value of each pixel as shown in Equation 1. Find the pixel P with the highest priority value and the target block $\varphi_p$ to be repaired with the P point as the center.

#### 1.2.2 Best match block searching

The global scanning method is used for searching, and the best matching block with the highest similarity to the target block $\varphi_p$ to be repaired is searched in the known sample area $\Phi$ by using the SSD matching criterion. The

definition of the SSD matching criteria is as follows:

$$SSD(p,q) = \arg\min_{\varphi_q \in \Phi} SSD(\varphi_p, \varphi_q) \quad (5)$$

$$SSD(\varphi_p, \varphi_q) = \sqrt[3]{\sum_{i=1}^{m}\sum_{j=1}^{n}\left[(p_{ij}^R - q_{ij}^R)^2 + (p_{ij}^G - q_{ij}^G)^2 + (p_{ij}^B - q_{ij}^B)^2\right]} \quad (6)$$

In the above formula, in order to calculate the similarity function between the known sample region and the target block to be repaired, where m and n respectively represent the length and width of the sample block, and R, G and B are respectively the known sample region $\varphi_q$ and the target block $\varphi_p$ to be repaired. The RGB components of the block of pixels. The best match block $\varphi_q$ is finally obtained.

1.2.3 Boundary update

The pixel value of the center point q of the best matching block $\varphi_q$ obtained in the above step is assigned to the target pixel point p to be repaired and the image is updated.

1.2.4 Boundary confidence update

Modify the confidence of the target pixel P to be repaired and update the entire image confidence. Repeat the above three steps until the entire image is repaired, which means $\Omega$ is empty.

1.3 Criminisi algorithm shortcoming

The Criminisi algorithm is an image restoration algorithm based on the sample region. The priority of the pixel P on the boundary to be repaired determines the order of image restoration. The value is determined by the product of confidence term C(p) and the data item D(p). In the actual reparation, as the repair process continues, the confidence of the repaired pixel needs to be updated which consults the confidence of the subsequently repaired pixel decreasing, resulting in the value being too low to lose credibility and reducing the image restoration effect. Moreover, when the best pixel P to be repaired is repaired during the repair process, the confidence of the neighboring pixels is increased and the priority of the neighboring pixels is increased, so that the pixel becomes the next best pixel to be repaired. The possibility of an increase in image repair order may result in continuity and a deviation in the image repair order affects the repair result. The selection of the best matching block $\varphi_q$ in the algorithm determines the matching information of the pixel to be repaired, and the selection is determined by the difference value of the RGB components of the pixel block to be repaired and the matching block. In the actual selection, the selection of the best matching block adopts the global search method, which causes the image with large RGB component to be easily matched to the pixel value whose texture does not conform to the visual effect, and the effect of the image restoration caused by the persistence affects the image restoration.

2 Algorithm in this paper

In order to obtain better repairing effect and improve the quality of image restoration, the Criminisi algorithm is used to redefine the confidence calculation method and introduce the Manhattan distance into the priority calculation because of the gradual failure of the confidence in the priority calculation formula. In the item, according to the known sample area and the boundary of the area to be repaired, the other pixel points in the target block to be repaired with the repair point are different to the central pixel point due to the influence of the surrounding condition distribution and the Manhattan distance in order to get the best pixel to be repaired. At the same time, for the problem that the best matching block is deviated due to the pixel difference of the corresponding pixel of the target block and the search matching block in the original Criminisi algorithm, the Euclidean distance is combined with the matching block pixel variance to find the best matching block. Therefore, the search is performed in a range adjacent to the feature as much as possible within a range acceptable to the pixel difference, and the accuracy of the matching block is improved to some extent. The algorithm flow chart of the algorithm is shown in Figure 2. The improved Criminisi algorithm optimizes both the confidence calculation and the best matching block search, and improves the visual effect after image restoration.

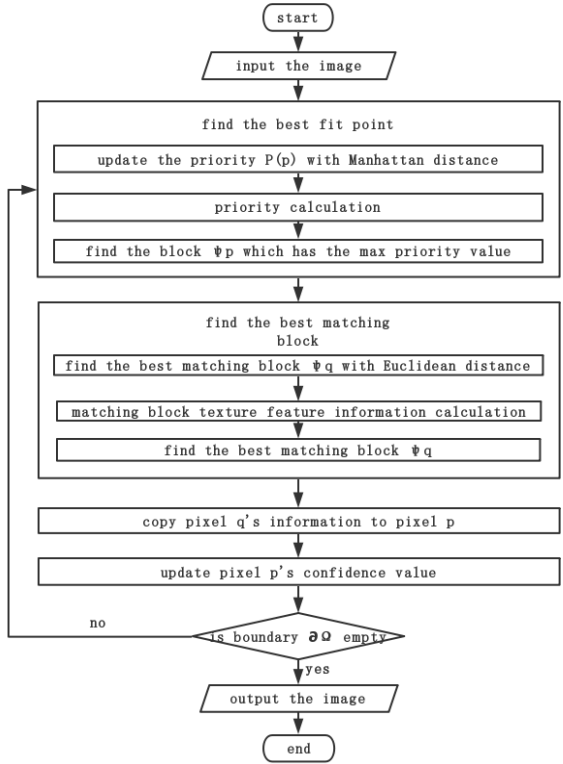

Fig. 2  Algorithm flow chart in this paper

2.1  Priority improvement

The original Criminisi algorithm has a tendency to rapidly decline and tend to zero before the digital image is not completed due to confidence, which leads to continuity of image restoration and disordered repair sequence. Therefore, the calculation formula for the confidence term in the improved priority calculation formula is as follows:

$$C(\hat{p}) = \frac{\sum_{q \in \varphi_p \cap \Phi} \frac{Dis(p,q)}{2} * C(p)}{|\varphi_p|} \quad (7)$$

$$Dis(p,q) = |p_i - q_i| + |p_j - q_j| \quad (8)$$

In the above formula, $\Phi$ denotes a known sample region, p is a target pixel to be repaired on the boundary, $\varphi_p$ is a target block to be repair centered on point p, and q is any pixel in the target block to be repaired. Dis(p, q) represents the Manhattan distance between two pixel points of p and q, where $p_i$, $p_j$, $q_i$, and $q_j$ are plane coordinates of two pixel points of p and q.

When the repair area become smaller, the confidence of the known sample area and the pixel to be repaired on the boundary of the target area to be repaired is reduced due to the reduction of the confidence of the pixel in the target block to be repaired. And because the confidence of the critical pixel will increase with the update of the pixel's confidence level, the repair region may continue to cause the repair sequence to deviate and affect the image's repair effect. The improved Criminisi algorithm proposed in this paper modifies the calculation of C(p) confidence terms in the priority calculation, and introduces the calculation method of combining Manhattan distance and pixel point confidence to change the repair order of the target area to be repaired, preventing the occurrence of the original Criminisi algorithm. Continuous repairs tend to reduce the confidence of the pixel points to be repaired. The improved C(p) confidence term introduces the Manhattan distance, so that the boundary pixel to be repaired calculates the priority of the pixel to be repaired due to the degree of its surrounded by known pixel points, which slows the decline of confidence and achieves the desired repair effect.

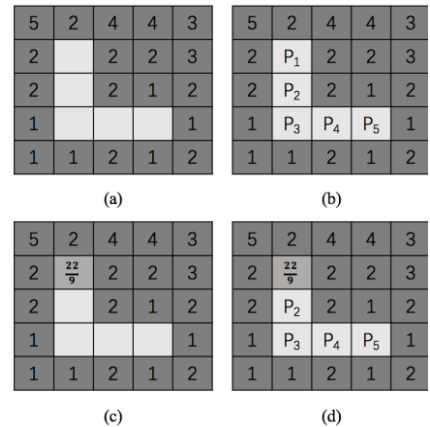

Fig. 3  Confidence distribution and updating schematic diagram

As shown in Figure 3(a), the confidence of the pixel and the position of the known sample of the pixel are defined as shown in the figure. The figure shows the unnormalized pixel point confidence of the known sample area. The blank space is the pixel to be repaired. As shown in Figure 3(b), the pixel number to be repaired is defined. Assuming that the size of the pixel block to be repaired is 3*3, it can be seen that C($p_1$) = (5 + 2 + 4 + 2 + 2 + 2) / 9 = 22/ 9, the same reason can be obtained C ($p_2$) = 1, C ($p_3$) = 1, C ($p_4$) = 7 / 9, C ($p_5$) = 11 / 9. According to the calculation method of the priority in the original Criminisi algorithm, C(p) is directly determined by the confidence degree, then the best pixel to be repaired tends to be $p_1$ in Figure 3(b), and if the repair confidence of $p_1$ is updated to 22/ 9, as shown in Figure 3 (c). Continue to perform the image repair step. Similarly, the confidence level of the remaining pixels to be repaired is updated to C($p_2$) = 103/81, C($p_3$) = 1, C($p_4$) = 7/9, C($p_5$) = 11/ 9, then in Figure 3 (d) the best pixel to be repaired tends to be $p_2$. then the repair order is biased towards $p_1$, $p_2$...

| 1 | 0.5 | 1 | $P_1$ | $P_2$ | $P_3$ |
| 0.5 | 0 | 0.5 | $P_4$ | P | $P_5$ |
| 1 | 0.5 | 1 | $P_6$ | $P_7$ | $P_8$ |

(a)          (b)

Fig. 4    The weight distribution in Manhattan distance

Using the improved Criminisi algorithm proposed in this paper, the confidence term of the pixel to be repaired is calculated by combining the Manhattan distance. As shown in Figure 4(a), the size of the pixel block to be repaired is also 3*3, and the P point is taken as half of the Manhattan distance from the other pixel points in the center pixel block to the P point in the center is its pixel point confidence weight. The confidence weight of the pixel presents different weights due to the difference in the distribution surrounding state of the pixel in the pixel block to be repaired. In the image repair process, the confidence of the pixel to be repaired is gradually updated during the repair process, resulting in a decrease in pixel value confidence. That is, in Figure 4(b), the point $p_1$, $p_3$, $p_6$, $p_8$ of the target pixel block to be repaired with the pixel point P as the center of the pixel to be repaired is higher than the reliability of the point $p_2$, $p_4$, $p_5$, $p_7$ of the point P, so the center pixel to be repaired is used in this paper. The half of the other pixel point Manhattan distance value is used as the weight to determine the confidence term, which appears as a diffuse increase trend. One is because the repair sequence has a tendency to repair along the Manhattan route during the image restoration process, and the other is because the half of Manhattan's distance value is slightly less than the Euclidean distance, which balances the confidence and weight to prevent the cause. If the distance is too large, it will affect the image repair order. According to the algorithm proposed in this paper, it can be seen that in Figure 3(a), C($\hat{p}_1$) = 5 * 1 + 2 * 0.5 + 4 * 1 + 2 * 0.5 + 2 * 0.5 + 1 * 1 + 2 * 1 = 16, the same reason Available C($\hat{p}_2$) = 7, C($\hat{p}_3$) = 8, C($\hat{p}_4$) = 5, C($\hat{p}_5$) = 9.5. In the calculation of the confidence term in this paper, the best pixel to be repaired in Figure 3(b) also tends to $p_1$. If the repair confidence of $p_1$ is also updated, the confidence level is 22/9, as shown in the Figure 3(c). Continue to perform the repair step of the improved algorithm proposed in this paper. The confidence of the remaining pixels to be repaired is similar to C($\hat{p}_2$) = 74/9, C($\hat{p}_3$) = 8, C($\hat{p}_4$) = 5, C($\hat{p}_5$) = 19/2, then in Figure 3(d) the optimal pixel to be repaired of the algorithm tends to be $p_5$, then the repair order is biased towards $p_1$, $p_5$...

By comparing the original Criminisi algorithm to tend to repair the repaired pixels in order (along $p_1$, $p_2$... trend), the improved algorithm proposed in this paper tends to select the more reliable pixel order of the confidence to repair, not only It can slow down the original Criminisi algorithm, which has a tendency to decrease the confidence quickly until the trend toward zero, which leads to a decrease in image accuracy. It also can reduce the error in the image repair order, so that the image repair effect is reduced.

## 2.2 Finding matching block method improvements

The Criminisi algorithm uses SSD function to find the optimal matching block. The SSD function only consider the RGB different between the sample block and the matching block, which is suit for the images with simple texture and color. However, it is extremely prone to matching problem, which causes image repair error, for the colorful images. Aiming at this defect, this paper introduces the Euclidean distance to improve the finding matching block method, which can fairly reduce the matching error. Image restoration usually repairs damaged areas based on neighborhood information of the area to be repaired. When finding matching block, the matching block search is performed in the neighborhood area, which can improve the image restoration effect. Therefore, the Euclidean distance between the sample block and the best matching block is also one of the important factors affecting the matching strategy. Figure 5 shows the image restoration effects at different search distances. The comparison matching block search area is limited to 60 pixels and 300 pixels. It can be seen that the repair effect is better when the search area is limited to 60 pixels.

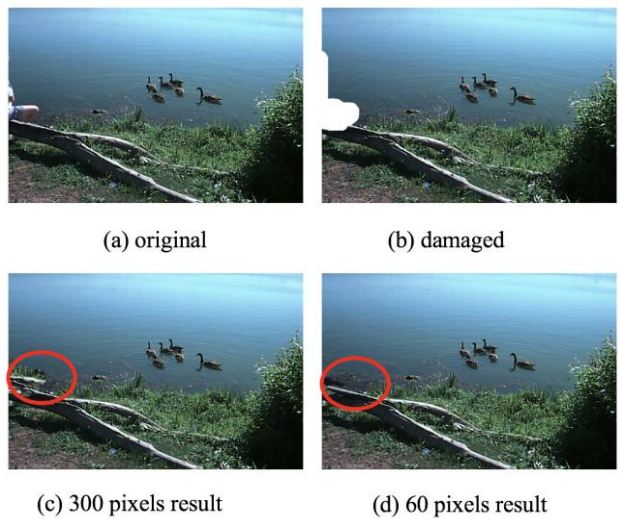

Fig. 5 Contrast figure based on the different search scope

D(p, q) is the Euclidean distance between the central pixel point p of the sample block and the central pixel point q of the best matching block, as shown in formula 9. Improve the matching strategy to formula 10, where m is the weight. Through the analysis of experimental data, the SSD value is different from the Euclidean distance D(p, q) by $10^2$ orders of magnitude. In the simulation experiment, m is between 0.009 and 0.01, and the better image restoration effect can be obtained.

$$D(p,q) = \sqrt{(p_i - q_i)^2 + (p_j - q_j)^2} \quad (9)$$

$$S\hat{S}D(\varphi_p, \varphi_q) = SSD(\varphi_p, \varphi_q) * m + D(p,q) \quad (10)$$

## 3 Experimental results and analysis

The experiment uses Microsoft Visual Studio Community 2015 as the simulation platform and is implemented in language C++. The hardware environment is 2.6 GHz Intel Core i7 processor and 12G memory. For the [2], [12] and the algorithm mentioned in this paper, the following several different scene images are selected for simulation experiments. Figure 6 to Figure 7 are experiments for repairing target removal, and Figure 8 to 10 are experiments for repairing damaged areas. The objective evaluation criteria were also used in the experiment. The peak signal-to-noise ratio (PSNR) and structural similarity (SSIM) were used to compare the image restoration effects. Tables 1 and 2 are the comparison results of the above three algorithms.

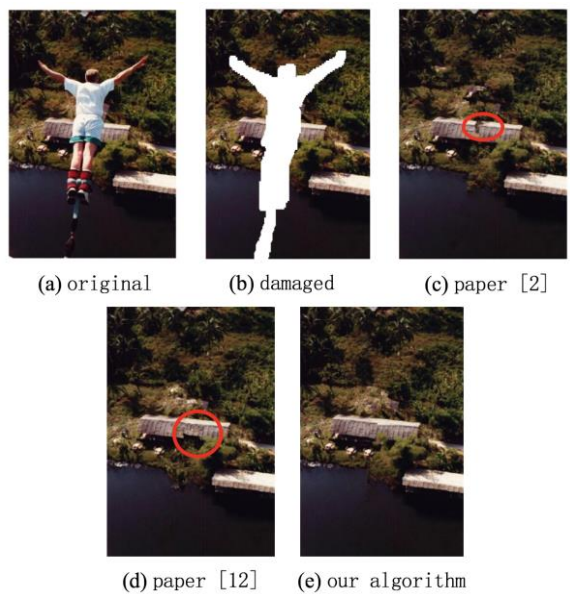

Fig. 6 Jump Result

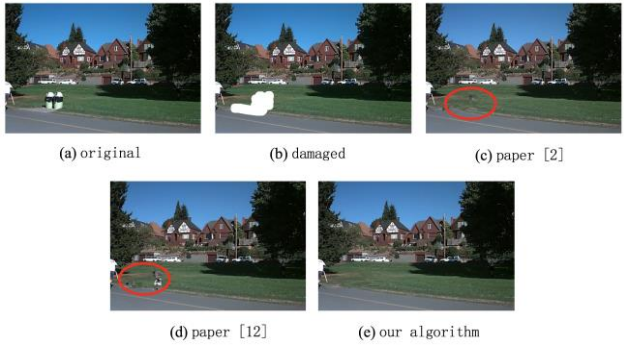

Fig. 7　Trash Result

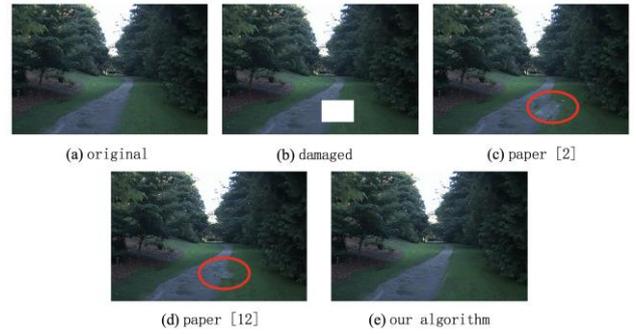

Fig. 9　Road Result

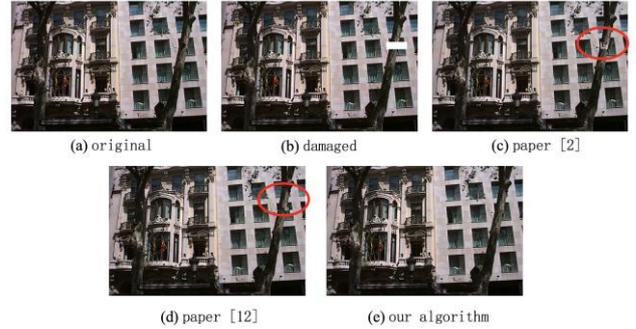

Fig. 10　Window Result

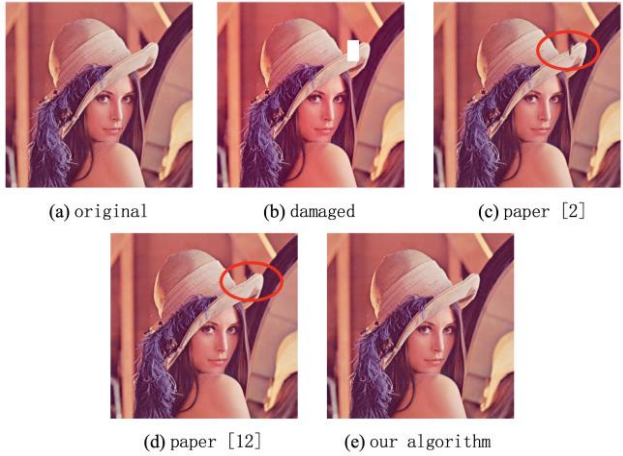

Fig. 8　Lena Result

Tab.1 Comparison of three methods' PSNR value

| Object | Size/Pixel | Paper [7] | Paper [12] | Our Algorithm |
|---|---|---|---|---|
| Lena | 512*512 | 38.8785 | 41.8085 | 42.1535 |
| Road | 756*504 | 34.9254 | 40.0016 | 42.6448 |
| Window | 756*504 | 33.3458 | 35.6003 | 40.0046 |

Tab.2 Comparison of three methods' SSIM value

| Object | Size/Pixel | Paper [7] | Paper [12] | Our Algorithm |
|---|---|---|---|---|
| Lena | 512*512 | 0.9984 | 0.9989 | 0.9989 |
| Road | 756*504 | 0.9883 | 0.9929 | 0.9941 |
| Window | 756*504 | 0.9923 | 0.9926 | 0.9937 |

From the results of Table 1 and Table 2, it can be seen that compared with the peak signal-to-noise ratio and structural similarity of [2] and [12], the improved Criminisi algorithm proposed in this paper obtains higher values and better repair effect.

In Figure 6(c) and Figure 6(d), the roof has defects due to the continuity error in the repair sequence. However, in Figure 6(e), due to the introduction of Manhattan distance weight, the image restoration has no continuity error. A repair image with better visual effects is obtained.

Both Figure 7(c) and Figure 8(d) result in deviations in image matching block selection due to single structural feature matching, and white breakpoints in the lawn affect the repair results. In Figure 7(e), the matching block selection algorithm which introduces the Euclidean distance combined with the pixel difference makes the image restoration texture feature smooth and achieves better results. In the repair results of Figure 8(c) and Figure 8(d), the texture features of the filled area are inconsistent due to the influence of the texture structure at background and the hat edge. The repair of the two different texture feature in Figure 8(e) has achieved good results.

The repair result of Figure 9(c) is quite different from the original texture. The better experimental results are obtained in Figure 9(d), but the continuity of the repair results is deviated, and the repair result of Figure 9(e) makes the edge connection more natural. In Figure 10(c), the color of the trees and the windows appear mixed. Figure 10(d) still shows the problem of continuity deviation of the repair results, while the repair effect of Figure 10(e) makes the trees and windows smoother and traceless.

By introducing the Manhattan distance as the confidence weight in the priority calculation formula of the original Criminisi algorithm, the image restoration can be repaired in the order of the pixel features around the pixel, preventing the image from being deviated due to the continuous repair. Secondly, the combination of Euclidean distance and pixel difference is added in the process of selecting the best matching block, and a better ratio between the distance and the pixel difference is obtained, so that the filled pixel is more natural and conforms to the surrounding image features. It can be seen from the above that the algorithm has improved the image restoration order and the search method, which makes the visual effect of using this algorithm better than other algorithms.

## 4  Conclusion

Through the research of the Criminisi algorithm, this paper analyzes its shortcomings both in the priority calculation and matching strategy. By studying the impact of the confidence update on the repair order, the priority calculation method is improved, and the repair error is also solved. By considering the locality of image restoration, the Euclidean distance is introduced to correct the matching strategy, which makes the matching block search more in line with the visual effect and improves the quality of image restoration. The experimental results show that the proposed algorithm can effectively guarantee the edge structure and texture information of the image, and make the repair effect more natural. The shortcoming of the algorithm in this paper is that because the algorithm still adopts the global repair strategy, the time complexity of the original algorithm is not optimized, resulting in long repair time. In the matching strategy, the determination of the weight lacks adaptability. The next step is to mine the image information, so that the weights in the matching strategy are more accurate, so as to further improve the repair effect. At the same time, for the global search strategy, the local search strategy can be used to improve the time complexity of the algorithm. How to accurately determine the scope of the search is still the focus of the next step.